# ACCESS CONTROL MECHANISMS FOR SEMANTIC WEB SERVICES - A DISCUSSION ON REQUIREMENTS AND FUTURE DIRECTIONS

*MANDEEP KAUR GONDARA*

Ph. D Student, Computer Science Department, University of Pune, Pune

*u08401@cs.unipune.ac.in*

**ABSTRACT:** *Semantic Web is an open, distributed, and dynamic environment where access to resources cannot be controlled in a safe manner unless the access decision takes into account during discovery of web services. Security becomes the crucial factor for the adoption of the semantic based web services. An access control means that the users must fulfill certain conditions in order to gain access over web services. Access control is important in both perspectives i.e. legal and security point of view. This paper discusses important requirements for effective access control in semantic web services which have been extracted from the literature surveyed. I have also discussed open research issues in this context, focusing on access control policies and models in this paper.*

## 1. INTRODUCTION
### 1.1 Semantic Web
The semantic Web is an extension of the current web in which information is given a well defined meaning, better enabling computers and people to work in co-operation [2]. It is the idea of having data on Web defined and linked in a way that it can be used for more effective discovery, automation, integration and reuse across various applications. The semantic web is an information space in which the information is expressed in a special machine-targeted language [1]. The evolution of semantic web will bring structure to the meaningful content of web pages.

### 1.2 Web Services
Software programs that can be accessed and executed via the web provide "web services". A service can consist of plain information, for example a weather forecast, or it may have an effect in the real world, for instance when booking a flight, ordering a book or transferring money[3]. Thus web services turn the 'static web' into a 'web of action' and bring the computer back as a device for computation.

### 1.3 Ontologies in Semantic web
Semantic Web techniques, particularly ontologies, facilitate portraying Web services with machine understandable semantics. Ontologies providing new features namely automatic composition, simulation and discovery of Web services [7]. An ontology is a formal, explicit specification of a shared conceptualization [6]. It is a system of concepts and their interrelationship between them. Ontologies are essential for semantic interoperability and advanced information processing as web services allow computation over the web. Semantic web services facilitate the dynamic discovery of services on the basis of a formal, explicit specification of the requester needs [9]. As an important component of the Semantic Web, ontologies will bring structure to the meaningful contents of web pages [8].

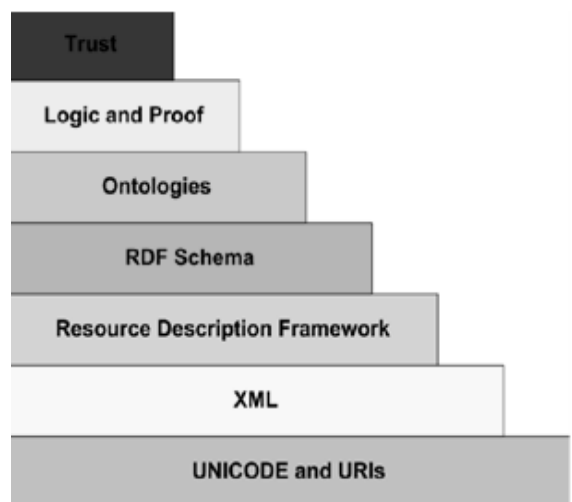

Figure 1. Semantic Web Layered Architecture

### 1.4 Security Challenges in Semantic Web
The Semantic web creates new security challenges due to its completely decentralized nature, the meta data description, the extremely large number of users, agents, and services. Security challenges associated with the semantic web involves the ability to handle security and to automate security mechanism to a more autonomous system that support complex and dynamic relationships between data, clients and service providers[10].

### 1.5 Policies and Semantic Web Services
Policies, which constrain the behavior of web services, are becoming an increasingly popular approach to Semantic Web. Policies are pervasive in web applications. They play an important role in enhancing security, privacy and usability of distributed services and indeed may determine the success or failure of a web service [4]. However users will not be able to personalize policies applied in the contexts. For web services, this includes policies for access control, privacy and business rules, among others.

Web service providers specify access control policies to restrict access to their web services. However, Internet is the open and distributed and dynamic environment, in which a central controlling instance cannot be assumed but still policies are needed to protect any system open to the internet [11].

### 1.6 Access Control in Secured Semantic Web Services

Access control is a mechanism that allows owners of resources/services to define, manage and enforce access conditions applicable to each resource. A semantic aware access control mechanism should assure that only eligible users are authorized to be granted an access right and each eligible user must be able to access all the resources that s/he is authorized for [5].

Web service providers want to restrict access to their services to only eligible users and Users want to access Web services. In case of restricted access they must prove their eligibility for getting the access over web services.

### 2. STATE OF THE ART

Access control systems for protecting web resources along with credential based approaches for authenticating users have been studied in recent years. With the advent of semantic web, new security challenges were imposed on security systems.

*The following security frameworks for access control in web services have been considered:*

Qin and Atluri [8] proposed a concept-level access control model for the semantic web. The control mode is used to specify access authorizations based on concepts and their relationships. Access authorizations are defined in ontologies and enforced upon data instances annotated by the concepts. The model consists of concepts and their relationships, propagated policies, authorization conflicts resolution and a semantic access control language. Their concept-level model is especially suitable for the specification and administration of access control over semantically related web data under the Semantic Web. They have discussed some domain independent relationships among concepts and also elaborated that how these relationships affect propagation policies. Their access control model also handles the way, the user read request for a document. In future work, they can explore the complexity of propagation and conflict resolution.

Rafae Bhatti et al. [14] have presented X-RBAC policy specification framework for enforcing access control in dynamic XML-based web service. An X-RBAC system has been implemented as a Java application, and is based on a specification language that addresses specific security requirements of these Web services. Their framework can be used to specify and enforce RBAC policies for securing XML documents at conceptual, schema, instance as well as element levels, and allows dynamically capturing context information. In future, they can add support for a multi-domain environment in their work where policy authorizations may not be centrally located, but are distributed across several domains.

Agarwal el al.[15] have presented a capability access control system where a web service provider specifies the access requirements for his service in terms of required properties. For gaining access, users have to prove that they satisfy these properties. To do so, they need to present certificate chains that prove a delegation chain from a trusted certification authority to the required property. In their paper, they have shown, (1) how certification authorities and their certification policies can be modeled semantically (2) how Web service providers can specify and check the consistency of their access control policies and (3)how the end users can check automatically, whether they have access to a web service or not.

In future, they can extend their work by finding out trusted root in certificate authority (CA). Potential CAs can publish signed certification policies in order to establish trust in them. While composing web services and access control policies from those of its components, a web service provider needs to check whether governmental and self imposed laws are still met by those CAs trusted by the components.

Carminati, Ferrari and Thuraibgham [12] proposed a security framework that uses RDF for policy specification and enforcement (RDF-PSE).The framework utilizes the semantic richness of RDF for expressing security information and hence making policy specification. The framework is capable of automatically entailing all the authorizations implied by the application of the high level policies to a specific scenario.

Sudhir Agarwal et al.[16] have given their contribution to the proof and trust layer of the semantic Web Layer by incorporating two suitably established techniques, i.e. (DARPA - An agent markup language for services) DAML-S (for describing Web services with machine-processable semantics) and (Simple public key infrastructure/ Simple Distributed Security Infrastructure) SPKI/SDSI (for specifying authorization based access control). Their approach was on the lines of autonomous granting of access rights and decision making on the basis of independent trust structures. The framework permitted the specification of access control related and functionality related aspects in a unified frame way that is manageable and efficient. Consequently, the approach is useful in distinctive Web service based applications (client-server architecture) and also in peer to peer and agent based applications.

Ashraful Alam et al. [13] have presented and implemented a security framework to prevent security problems. They defined a modular access control policy framework in the perspective of geospatial data integration platforms. Geospatial semantic Web services used the framework to impose resource access and intelligently make decisions about policy rules. The framework permitted reasoning capabilities at both the resource enforcement point and service discovery point.

From the literature surveyed, it is evident that existing frameworks were developed to address different security aspects of the semantic web. It can be observed that most of the research work is being done for carrying out the semantic discovery of web services in a secure manner. But there is a need to develop a framework which would provide semantic web services that are relevant to the user request, and only to those users who have got the access rights.

## 3. SOURCES OF SURVEY

The requirements of access control mechanism in semantic web services have been extracted from the literature surveyed. In the research, I used documentary sources, which involve existing textual documents available in electronic and printed media. The data sources used in this research include academic journals, applicable books and the internet.

The requirements were extracted from existing access control models and mechanisms and other surveyed literature on access control aspects in the semantic web services.

Different authors have indicated different aspects that should be considered while designing an access control mechanism for the semantic web services. These aspects include many parameters as discussed in section 4 for access control mechanism in semantic based web services.

## 4. REQUIREMENTS FOR ACCESS CONTROL MECHANISM

An Access control mechanism can ensure that only eligible user can get access to a web service. The access control policy of a web service is specified by the provider of the web service description. An end user may know some web service may combine few of them in some way to solve a certain task at hand. Prior to executing such a combination or plan, the user must know whether he can fulfill the access control policy of the plan [16].

### 4.1 Access Control Mechanism should satisfy composite web services

Web services can be categorized into two types: atomic and composite web services. Atomic web service is a service that is complete in it and cannot be broken further. Composite web service is a service that can be broken into atomic and composite web services. The access control mechanism should satisfy composite web services not only atomic services. For e.g., a web service say w3 is dependent upon the other web services say w1 & w2. A web service w2 can be only executed after the completion of w1. The mechanism should support interplay of the access control in composite web service.

### 4.2 A Mechanism should satisfy the complex requirements of an access control

This requirement can be specified as access control policy of the Web service easily. However, the access control policies of most of the Web services are not so simple. An access control policy of web service may depend on the requested functionality and the provided functionality (for e.g., values of the input parameters). The functionality may depend on the access control conditions fulfilled by the user [16]. A mechanism or model should be able to satisfy the complex requirements of an access control of web services.

### 4. 3 Semantic relations among concepts should be considered

It is essential to consider the semantic relationships among concepts while designing an access control mechanism as semantic relationships between concepts play a key role in making access control decisions. Security can be violated if access control to each concept is considered separately ignoring the interrelationships among concepts. Information may be inaccessible to authorized subjects if relationships among concepts are not considered. An access control to the semantic web needs to additionally ensure that all the information authorized to be viewed should be revealed to a subject [8].

### 4.4 Incorporation of policies in Access Control

Access control policies are needed to protect any system open to the internet. Policies play crucial roles in enhancing security and privacy in semantic web environment. Policies make decisions based on properties of the peers interacting with the system. These properties may be strongly certified by cryptographic technique. For e.g., age, nationality, customer profile, identity, and reputation may all be considered both in access control decisions, and in determining which discounts are applicable (as well as other eligibility criteria) [4]. We view access control policies as conditions a Web service provider defines to restrict the number of users who may access the functionalities offered by his Web service.

### 4.5 Credentials Consideration in Access Control Model

The access control model must consider credentials that are issued to the user during the execution of the web service. Credentials state a binding between a user and some property. Credentials are digitally signed documents, which can be transmitted by untrusted channels like the Web [16]. If the user has provided some information to web service provider then the provider should not disclose this information to retain secrecy of users' credentials.

### 4.6 Authorization must be considered over authentication

Most of the existing works are based on authentication procedure for access control. However the semantic web is an open, distributed and dynamic environment where web services should be offered to the user dynamically and instantly. Many access control models allow users to get access over the web services by proving their identity through registration. However, it happens that a user does not want to disclose his identity but wants to access the web service. By using authorization policy, the user can get access over a web service without any kind of registration.

## 5. FUTURE DIRECTIONS & RESEARCH ISSUES

The parameters stated in section 4 can be considered while designing an access control mechanism for accessing semantic based web services. However, the researchers have worked upon the parameters stated in section 4 but still a mechanism is required for access control over web services that fulfills all the above stated parameters.

*Here are some future directions for research work in semantic based web services for effective access control:*

### 5.1 Development and Integration of Privacy and Access Control Policies

Access Control policies are needed to protect any system open to the internet. Privacy policies are needed to assist users while they are browsing the web and interacting with web services. Privacy policies are the policies what I have referred as the user credentials that has to be made confidential. It is appealing to integrate these policies into a coherent framework, so that (i) a common infrastructure can be used to support interoperability and decision making and (ii) the policies can be harmonized and synchronized [4]. For example, age, nationality, customer profile, and identity may all be considered in access control decision. A user wants that the service provider should not reveal this information to outside world may consider as a part of privacy policy.

### 5.2 Policy Matching Mechanism

A policy matching mechanism is to be developed and incorporated in the framework. The matching mechanism can obtain the user policies from the user who is requesting the web services and the service providers' features can be obtained from the ontology and matchmaker can finally perform the job of policy matching on the basis of integrated privacy and access control policies mechanism.

In matching mechanism, the access can be accomplished by developing matching mechanism for the service providers and the requesting user. The user could be well-analyzed with his/her credentials. Based on the credentials, the access mechanism can check whether the user is eligible to access all the identified services. Then, based on the service providers' policies, the mechanism can check to which extent the users have the access rights to the services.

### 5.3 Service Matching Mechanism

Like the aforesaid policy matching mechanism, a service matching mechanism is to be developed in parallel to detect the similarity between the requested service and the services available in the ontology in an effective manner. This service matching mechanism can ultimately perform the discovery of web services that are semantically matched with the user request. This service matching mechanism is to be incorporated in the framework along with the access control mechanism for better discovery of semantic based web services.

## 6. CONCLUSION

The semantic web creates new security challenges due to its completely decentralized nature. Various efforts have been made to develop access control mechanisms for discovery of semantic web services which have been elaborated in section 2. There are still some research issues that are to be addressed in access control mechanism for semantic web services. Establishing the requirements of an access control mechanism for semantic web services is a critical milestone in the development of a security framework for the semantic web. In this paper, requirements of an access control in semantic web environment were proposed. Secondly, future directions for research in the same area have been proposed.